# Living Structure + AI
# A New Episteme for Liberal Arts and Whole-Person Education in the Era of Artificial Intelligence

Bin Jiang
LivableCityLAB, Thrust of Urban Governance and Design
The Hong Kong University of Science and Technology (Guangzhou), China
Email: binjiang@hkust-gz.edu.cn

**Abstract**
The rapid integration of artificial intelligence (AI) into higher education marks a radical epistemic shift, unsettling long-held assumptions about how knowledge is made, taught, and judged. This paper argues that the theory of living structure offers a timely foundation for liberal arts and whole-person education in the era of AI. Living structure—Christopher Alexander's term for the recurrent hierarchical order that makes environments feel alive—is treated here as two faces of one phenomenon: living, the right-brain perception of wholeness, and structure, the left-brain order that can be computed through fifteen geometric properties, two fundamental laws, two design principles, and the computable L- and B-scores. From this basis the paper develops a Living Structure + AI paradigm, a new episteme for higher education, deliberately so ordered because living structure is held to be more fundamental than AI: structure is the order of nature and culture, while AI is a tool that gives that order new expression. The argument is grounded in teaching practice led by LivableCityLAB, under its own teaching-research project and in partnership with Residential College 1, spanning architecture and city-science courses or theses, an undergraduate whole-person common-core course (Self and Wholeness), and a new postgraduate course (Experiential Learning: Living Structure + AI Inspired Design). Across these settings, students read built environments structurally, declare a skeleton, and use AI as a structural mediator while making and inhabiting real spaces. Across four campus renovations (total N = 33 students) and the two courses (N = 49 undergraduate; N = 7 postgraduate), structural scores (the L- and B-scores) rose in every renovated case, and these architectural and perceptual measures converged with preference tests and visual-attention analysis; course-embedded, rubric-based assignments provide complementary, though not psychometrically validated, evidence of students' engagement with the paradigm. The paper closes by arguing that the deepest contribution of Living Structure + AI to liberal arts education is not on paper or on screens but down to earth—in the daily life spaces in which students learn to feel, measure, and remake wholeness.

Keywords: Liberal arts education, Whole-person education, Higher education, Epistemology, AI in education

## 1 Introduction

Artificial intelligence (AI) has arrived in education as both promise and threat. It can tutor, summarize, translate, and generate at a scale no teacher can match; it can also hollow out the very capacities a liberal education exists to cultivate—patient attention, independent judgement, the felt sense of what is true and what is beautiful. Nowhere is this tension sharper than in the question of what, in an age of automated content, education is still for. More than a new set of tools, AI marks a shift in the very episteme of higher education—in what counts as knowing, making, and judging—and presses educators to rethink inherited ontologies, epistemologies, and pedagogies. This paper proposes an answer drawn from an unexpected quarter: the theory of living structure, developed by Christopher Alexander (1979, 2002–2005) to explain why some buildings and places feel alive while others do not. In brief, living structure is Alexander's account of a recurring, hierarchical kind of order—visible in a well-loved street as much as in a leaf or a river—that makes an environment feel coherent and alive rather than flat and inert; a pattern language, the related device he developed earlier, is a shared vocabulary of recurring design solutions to recurring problems (a window seat that gathers light and conversation, a street cafe that draws life to a sidewalk) that a community of builders and users can hold in common and use to describe, teach, and reproduce that quality. The claim of this paper is that living structure, so understood, far from being a narrow concern of architects, supplies a robust and teachable foundation for liberal arts education in the era of AI. Alexander's ideas have, in fact, already crossed into education: his pattern-language method has been adapted into pedagogical and learning pattern languages (Bergin 2000; Iba et al. 2009), and the fifteen properties at the heart of his later theory of wholeness—recurring, recognisable qualities such as levels of scale, strong centers, thick boundaries, and gradients that Alexander found common to environments people experience as alive, and which are introduced in full, with all fifteen named, in Section 2 and Figure 2—have been mapped onto learning design (Baumgartner and Bergner 2015).

This paper addresses significant gap in literature. Prior work, in other words, has already carried Alexander's ideas some distance into education: pedagogical pattern languages exist, and the fifteen properties have been mapped onto learning design in the abstract. What remains largely unexplored is the measurable, computable side of living structure—the fifteen properties together with the L- and B-scores—bound to AI and embedded in whole-person higher education as lived, hands-on practice, rather than as a purely conceptual mapping; that gap—not whether the fifteen properties can be described for educators, which prior work has already done, but whether they can be measured, taught, and put into students' hands together with AI as a working part of a liberal arts curriculum—is what this paper addresses.

The paper's main argument, stated at the outset, is this: because living structure can be simultaneously felt by the right brain and computed by the left, it offers a rare, teachable common ground on which liberal arts and whole-person education can reunite perception and analysis in the era of AI—provided AI is bound to living-structure criteria rather than left to set its own. Three motivations frame this argument.

The first motivation begins with a crisis in the built environment. Much of what has been built over the past century is, on the evidence of public preference and a growing body of research, experienced as flat, alienating, and lacking in life (Jacobs 1961; Salingaros 2006; Curl 2018; Boys Smith 2016; Heatherwick 2023). The built-environment claims made in this paper are deliberately scoped to this everyday, non-monumental territory—the classrooms, dormitories, and activity centres that make up the daily fabric of a campus—rather than to architecture, or any particular architectural style, in general. The spaces in which students study, sleep, and gather—classrooms, dormitories, activity centers—are too often visually undifferentiated and perceptually impoverished. This is not merely an aesthetic complaint. A substantial literature links environmental quality to health, recovery, and well-being (Ulrich 1984; Kaplan and Kaplan 1989; Wilson 1984; Salingaros 2015; Mehaffy and Salingaros 2020). The crisis of architecture is therefore also a crisis of perception: surrounded by environments that suppress the felt qualities of wholeness, students are gradually trained not to notice them. A liberal arts education that overlooks its physical environment misses one of the most pervasive influences on human development.

The second motivation concerns AI itself. The dominant educational use of generative AI is content production—drafting, illustrating, and answering. Recent studies of how students actually incorporate these tools in higher education confirm the pattern: use is dominated by brainstorming, drafting, and editing, even as students remain wary of ceding judgement to the machine (Pedroni 2026). Used this way, AI tends to reinforce a purely left-brain, analytical mode of learning while leaving the right-brain capacities of intuition, feeling, and aesthetic perception underdeveloped. Yet recent work shows that large language models, properly prompted, can judge architectural and urban beauty in close agreement with public preference and with measured brain response (Boys Smith and Salingaros 2025; Jiang 2025, Lavdas et al. 2021, Sussman and Hollander 2014). This suggests a different role for AI in education: not as a machine that produces content on our behalf, but as a cognitive amplifier that helps us see more acutely—to perceive levels of scale in a façade, order in nature, wisdom in tradition, and wholeness in ourselves. The educational question becomes how to design learning so that AI strengthens rather than supplants human perception. This echoes a long-standing question about technology and human capability: whether technology will enhance human skill, judgement, and creativity, or reduce people to minders of machines—architect or bee (Cooley 1987). Generative AI puts that question to education with renewed force.

The third motivation is that liberal arts and whole-person education are, at root, an attempt to develop the whole person—to reunite rational thought with sensible perception, knowing with feeling and being. Living structure is unusually well suited to this aim because it is intrinsically two-sided: it can be felt by the right brain as life and wholeness, and it can be analyzed by the left brain as computable structure (Jiang 2019). It thus offers a rare common ground on which science and art, measurement and meaning, can meet (Snow 1961, Brockman 1996). And because living structure is found not on paper but in real rivers, streets, rooms, and bodies, it grounds education in lived space (Dewey 1938). The ambition of this paper, echoing a familiar phrase, is that students learn to design and construct not only on screens and in journals but down to earth—in the daily life spaces they can touch, measure, and make more alive (Jiang 2019). Pursued seriously, this becomes a question for higher education in the age of AI as well: what habits of attention, judgement, and care a generation should carry into a world in which machines increasingly make and mediate the spaces, images, and texts of everyday life.

This paper makes four contributions. First, it reframes living structure—usually discussed within architecture and urban science—as a general foundation for liberal arts and whole-person education, showing that a single, teachable phenomenon can engage both hemispheres of the brain at once. Second, it articulates a Living Structure + AI paradigm and defends the ordering of its terms, positioning AI as a structural mediator and cognitive amplifier bound to scientific criteria of beauty rather than as an autonomous content machine. Third, it reports concrete, multi-level teaching practice led by LivableCityLAB—spanning architecture and city-science courses, an undergraduate common-core course (Self and Wholeness), and a new postgraduate course—through which the theory is turned into perceivable, experiential, and verifiable practice. Fourth, it draws out the implications of this approach for liberal arts education and for the transformation of design teaching, insisting that learning move from paper and screens down to earth—to the daily life spaces students can touch, measure, and remake. Across the four runs a single central message: in the era of AI, the most consequential act in education and design alike is the choice of criteria; living structure supplies criteria that are humane, teachable, and partly computable, and AI becomes socially beneficial precisely when it is bound to them.

The remainder of the paper proceeds as follows. Section 2 sets out the theory of living structure: its three dimensions, fifteen properties, two laws, two principles, and quantitative indicators. Section 3 develops the Living Structure + AI paradigm and explains why the ordering is deliberate. Section 4 sets out the research design, participants, and measurement instruments used across all reported settings. Section 5 describes teaching practice across three settings led by LivableCityLAB. Section 6 discusses the implications for liberal arts education and for the transformation of architecture and city-science courses. Section 7 concludes and points to future work.

## 2 Living Structure: A Theory of Wholeness Felt and Computed

Living structure refers to the order that makes environments feel alive—coherent, welcoming, and deeply satisfying. Articulated by Alexander (2002–2005) across the four volumes of The Nature of Order, building on the earlier pattern language (Alexander et al. 1977), and increasingly described and measured by subsequent work (Salingaros 2006; Jiang 2015; Jiang and de Rijke 2023; Jiang 2025), the theory begins from a simple observation: the difference between places that feel alive and places that do not is not arbitrary. It can be described, partly explained, and—crucially for education—taught. Living structure is found in nature, in vernacular settlements, and in well-loved cities (Mandelbrot 1982; Bak 1996; Eglash 1999), and it can be deliberately produced, although contemporary practice often obstructs it. To keep this vocabulary consistent throughout the paper: living structure names the theory and phenomenon itself; the fifteen properties, the two laws, and the two design principles introduced below are its descriptive and generative vocabulary; the L-score and B-score are the two computable, statistical instruments used to measure it; and Living Structure + AI, introduced in Section 3, names specifically the two-stage design-and-teaching paradigm built on top of this theory—not the theory itself.

The three dimensions of living structure. Living structure has three complementary dimensions, organized here around the two hemispheres of the brain (Figure 1). Living is the emotional, sensible, right-brain perception of an environment as whole, alive, and resonant with one's own sense of self—creativity, intuition, feeling. Structure is the computable, measurable, left-brain face of the same phenomenon—analysis, logic, hierarchy, and quantitative indicators. Living structure, the third dimension, is the unity of the two as engaged by the whole brain: the felt and the computed treated as one phenomenon seen from two sides. This left-brain/right-brain contrast is used, following McGilchrist (2009) and the broader literature on hemispheric specialization in attention (e.g., Gazzaniga 2000), as an organizing heuristic rather than a claim of strict neuroanatomical localization; what matters for education is the reality of the two modes, not their precise seat in the brain. The design and educational program developed below rests on bringing the two into systematic correspondence: what the left brain can compute helps us design for what the right brain can feel.

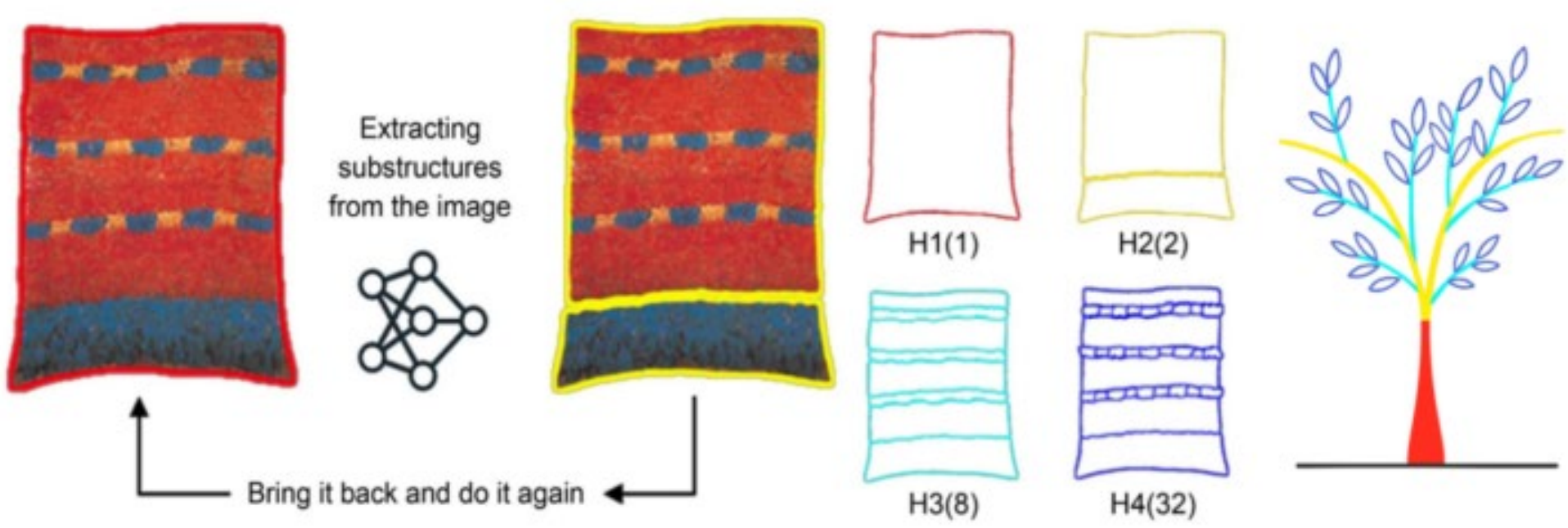

Figure 1: (Color online) Three dimensions of living structure, organized around the two hemispheres of the brain: structure as the left-brain face (analysis, logic, facts, hierarchy), living as the right-brain face (creativity, intuition, feeling, imagination), and living structure as the unity of the two engaged by the whole brain.

The fifteen properties, and their educational relevance. Alexander (2002–2005) identified fifteen geometric properties that recur in environments experienced as alive—a vocabulary distilled from long observation of vernacular architecture, sacred buildings, settlements, and natural systems. The fifteen are: levels of scale, strong centers, thick boundaries, alternating repetition, positive space, good shape, local symmetries, deep interlock and ambiguity, contrast, gradients, roughness, echoes, the void, simplicity and inner calm, and not-separateness (Figure 2). They can be grouped by importance and by the scales at which they operate: a small number of global properties span all scales, others govern local relations among substructures, and strong centers and good shape act as consolidating properties that the rest tend to produce. Levels of scale and not-separateness, for example, act globally across an entire form; alternating repetition, positive space, and local symmetries govern the relations among neighboring parts; and strong centers and good shape consolidate the whole. Together the fifteen amount to a generative grammar of life in space, equally applicable to a column, a wall, a room, a garden, a street, or a flower arrangement—and, by extension, to the structure of a life.

Although Alexander developed the fifteen properties for buildings and cities, this paper deliberately transfers them, in Sections 5 and 6, to two further objects of study: learning spaces themselves (Section 5.1), and the reflective vocabulary students use to describe their own coherence and growth—the self as a living structure theme running through Self and Wholeness (Section 5.2) and the postgraduate course (Section 5.3). The properties are not offered here as a general theory of pedagogy; rather, they function as a shared, teachable vocabulary that lets a design judgement about a wall or a room, and a self-reflective judgement about one's own habits or growth, be discussed in the same terms. This double use—of the same fifteen-item vocabulary for buildings and for selves—is the paper's central pedagogical device, and it is what is meant, throughout Sections 5 and 6, by treating living structure as a foundation for whole-person education rather than only as an architectural theory.

The two laws. Two fundamental laws underlie the properties. The scaling law (Jiang 2015) states that any coherent, living whole exhibits far more small substructures than large ones—visible in the branching of rivers, the venation of leaves, and the range of room sizes in a beloved house. Tobler's law (Tobler 1970) states that nearby centers or substructures tend to be similar in size and character, giving well-formed environments their characteristic local rhythm. The two laws are complementary: scaling governs hierarchy across levels; Tobler governs similarity within a level. In teaching, the pair is easy to demonstrate: a beloved street shows many small windows and openings and few large ones (scaling), while neighboring shopfronts stay close in size and rhythm (Tobler). Students quickly learn to see where a modern frontage breaks one law, the other, or both.

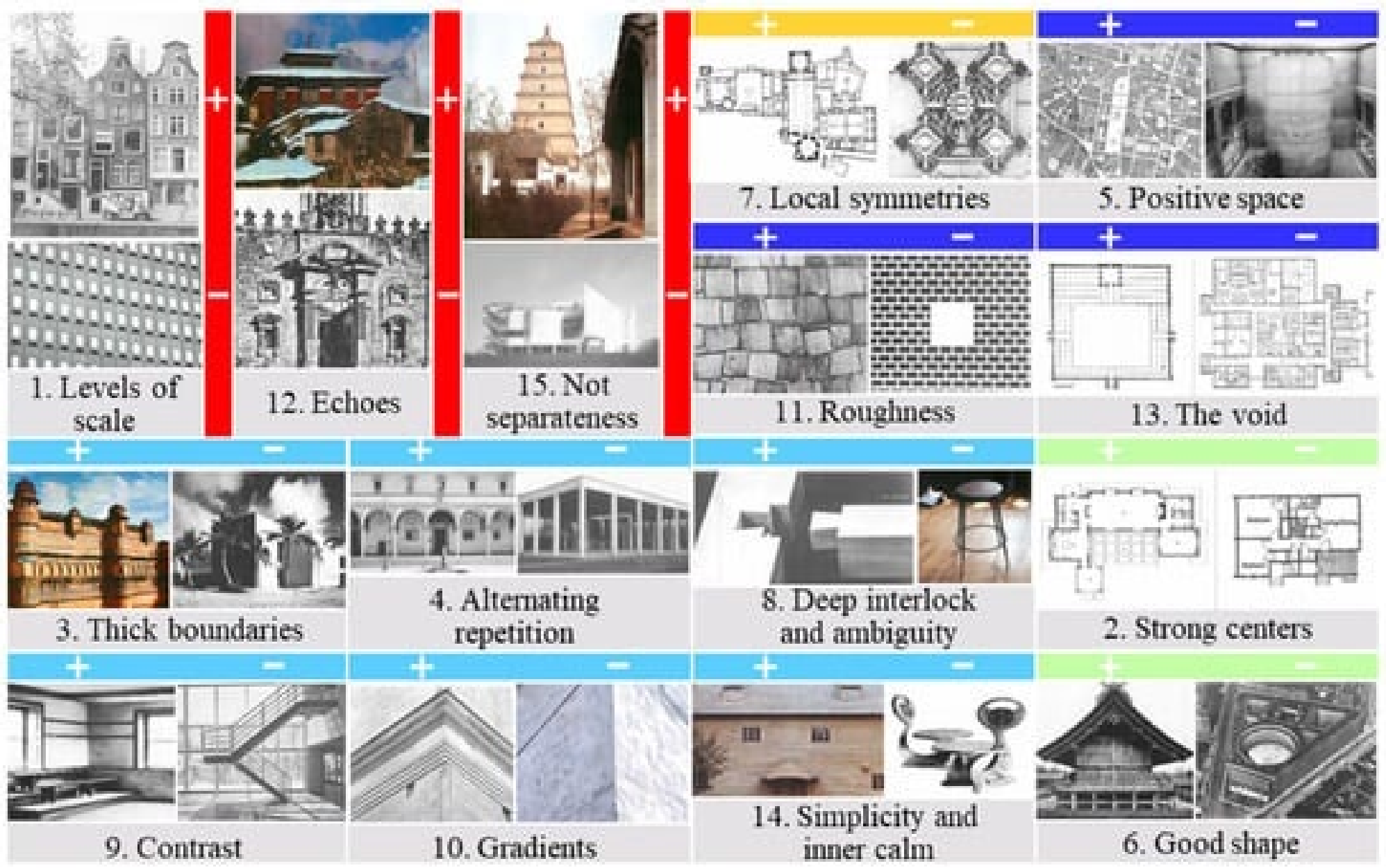

Figure 2. (Color online) The fifteen properties of living structure, organized into five groups by their relative importance and by the scales at which they operate. Three global properties (red) span all scales; local symmetries (orange) form a group of its own; two further groups (blue) capture local properties at single significant substructures and at sets of adaptive substructures; strong centers and good shape (green) act as consolidating properties that the others tend to produce.

The two design principles. Two design principles turn the laws into a generative procedure. Differentiation is the process by which an undifferentiated or less differentiated whole is gradually subdivided into coherent centers at multiple levels of scale. Adaptation is the process by which substructures adjust to one another and to their context (Alexander 2002–2005). Together, differentiation and adaptation convert living structure from a descriptive notion into something a student can actually do, step by step, with hands and tools.

The two computable indicators. Two indicators make the structure side measurable. The L-score (Jiang and de Rijke 2023) captures the depth and richness of hierarchical substructures, formalized as livingness equal to the number of substructures multiplied by their hierarchical levels ($L = S \times H$). Compared side by side, a richly subdivided traditional façade yields a far higher L-score than a sparsely subdivided modernist one, all else equal (Figure 3)—a comparison that reflects a broader, previously reported pattern across many buildings and styles (Xue and Jiang 2026), and which we discuss with an important caveat below. The B-score, produced by the Beautimeter (Jiang 2025), harnesses large language and vision models to rate images against the fifteen properties. Because the Beautimeter's scoring framework is itself derived from the same fifteen properties that define the L-score, agreement between the two indicators is best read as an internal consistency check—evidence that the two instruments measure the same construct consistently—rather than as independent, theory-external validation of living-structure theory or of the Beautimeter's judgements against some theory-free ground truth; we return to this distinction, and to what would be needed to move beyond it, in Section 4 and in the Limitations discussion of Section 7. Across thousands of images spanning many architectural styles, pre-industrial styles consistently score higher than modernist ones on average (Xue and Jiang 2026), and these structural scores align with public preference (Boys Smith and Salingaros 2025) and, in recent multimodal experiments, with measured brain response (Boys Smith and Salingaros 2025; Jiang 2025, Lavdas et al. 2021, Sussman and Hollander 2014). Importantly, as with temperature, more is not always better: a moderate, comfortable level of livingness, rather than a maximum, is itself a worthwhile research question. The indicators are instruments of conversation, not oracles—but, used in that spirit, they turn aesthetic intuition into something that can be discussed, taught, and refined. Three caveats deserve emphasis, and the paradigm absorbs rather than denies them. Whether beauty can be measured at all remains philosophically contested, and Alexander's theory still occupies a minority position within mainstream architectural discourse; aesthetic preference varies with culture, training, and context, and experts and lay publics often diverge; and large models inherit stylistic biases from their training data, so their judgements must be audited rather than presumed neutral. Hence the stance taken here: the scores are statistical and advisory, the criteria are public and revisable, and the final word belongs to human perception.

A further caveat concerns the façade comparison in Figure 3 specifically. Modernist buildings were not designed to be judged against Alexander's criteria; they embody their own coherent design philosophies, and Le Corbusier's own writings, for instance, defend a deliberate, reasoned rejection of applied ornament rather than an absence of design thought. A lower L-score is therefore not a verdict on such a building's intentions, historical significance, or overall architectural merit; it registers only one specific, narrow, computable property—the depth and multiplicity of hierarchical substructures that the L-score is defined to count—and not beauty, quality, or aliveness in any all-things-considered sense. Nor is pre-industrial ornament the only route to a high L-score: richly differentiated contemporary buildings, and non-Western modern buildings that use structure rather than applied decoration to create hierarchy, can and do score highly. The comparison is offered as an illustration of what the L-score computes, not as a global aesthetic ranking of modernism against tradition.

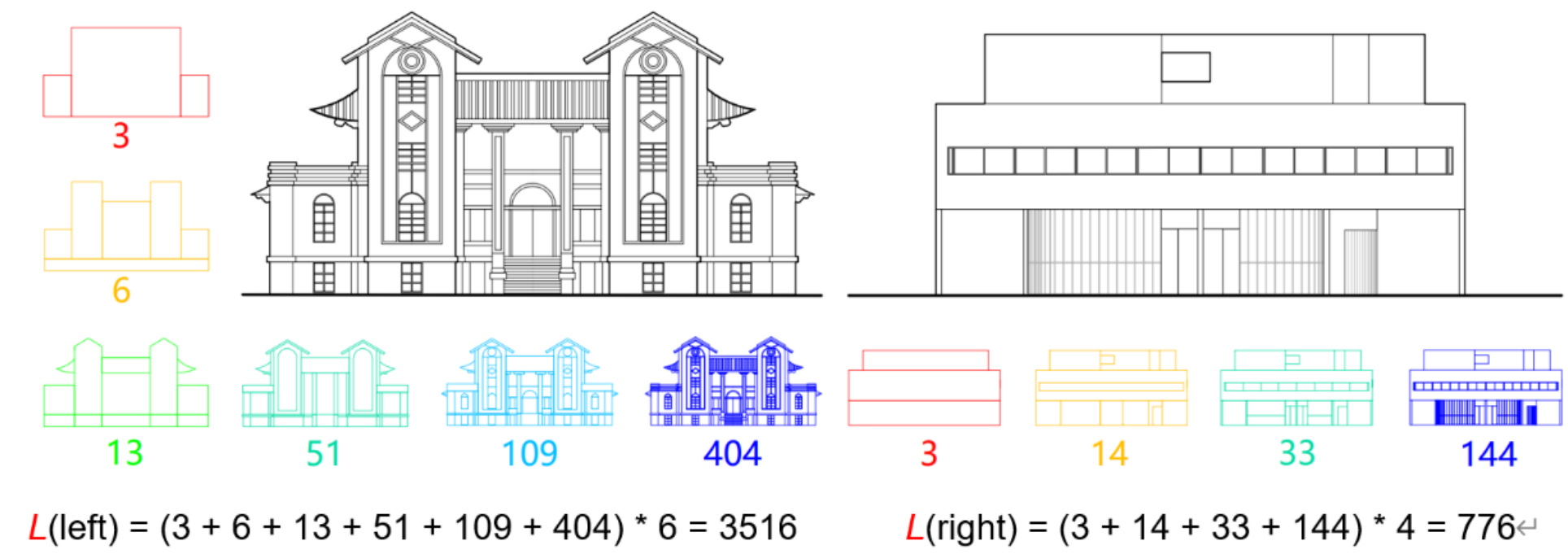

Figure 3: (Color online) Computing livingness as L = S × H (substructures × hierarchy) for a richly subdivided façade compared with a sparsely subdivided one. The comparison illustrates what the L-score computes—the depth and multiplicity of hierarchical substructures—and is not offered as an all-things-considered aesthetic verdict on any named building or architectural movement; see the discussion and caveat above.

Because the same order can be felt and measured, living structure is unusually teachable. A student can be shown how to count substructures and levels and, in the same breath, asked which of two façades feels more alive; the two judgements can then be laid side by side and argued about. This is also why the theory travels so naturally from buildings to the person. Alexander held that a living whole is one in which we recognise something of our own self, and the courses described below take this seriously, inviting students to read the self as a living structure—a hierarchy of centres that can be more or less coherent, more or less alive—and to treat the cultivation of inner wholeness as continuous with the making of beautiful space.

**3 The Living Structure + AI Paradigm**

The paradigm proposed here reframes design—and, as Section 5 shows, education—as a two-stage process: first the deliberate construction of a structural skeleton that satisfies living-structure principles, and then the AI-assisted generation of stylistic skins on that skeleton (Figure 4). The skeleton carries coherence, hierarchy, and life; the skin carries cultural and stylistic specificity. The two together yield results that are structurally coherent yet stylistically diverse.

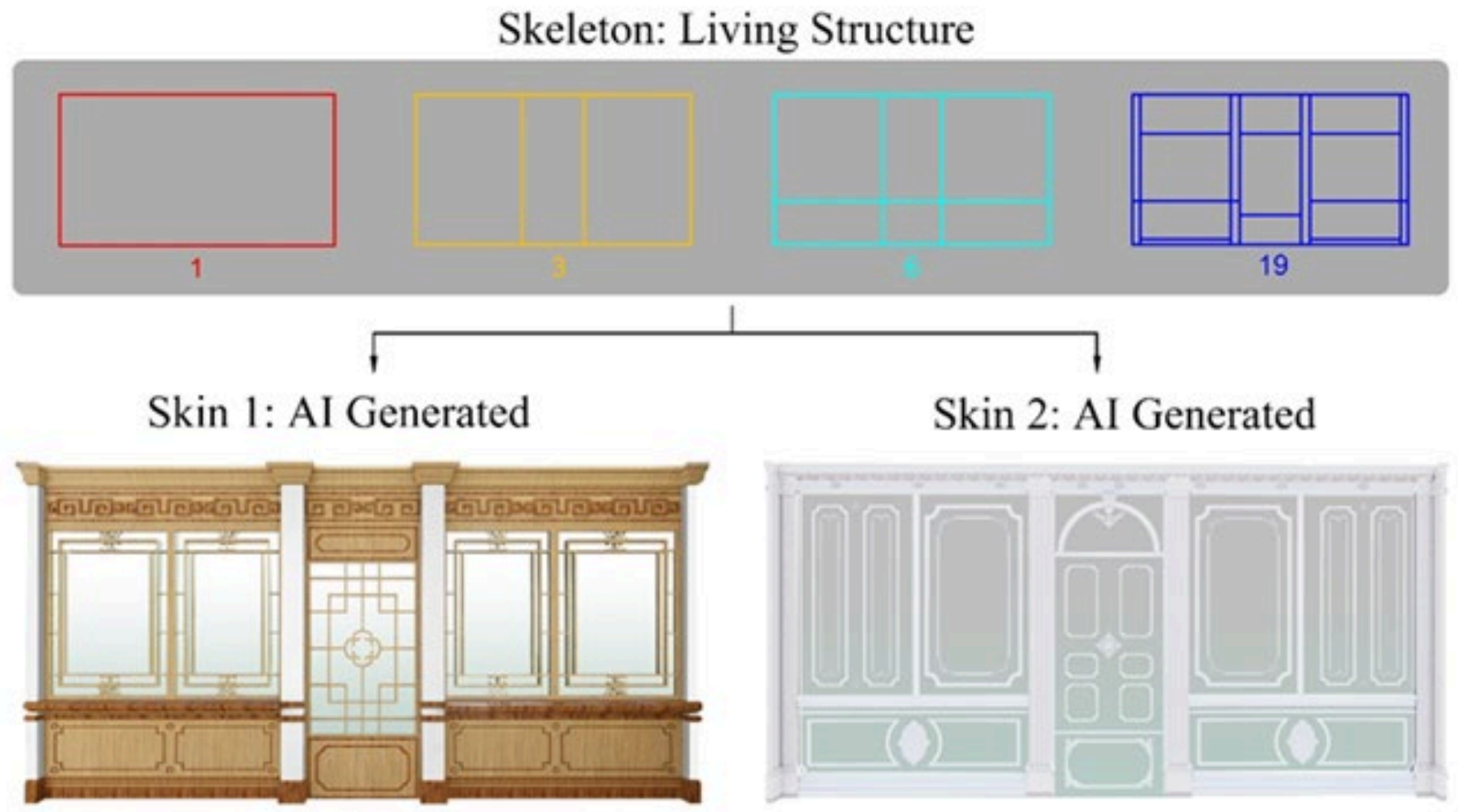

Figure 4: (Color online) The Living Structure + AI paradigm. A structural skeleton is built by iterative subdivision (1 → 3 → 6 → 19) so that there are far more smalls than larges; AI then generates alternative stylistic skins—here a contemporary Chinese and a more European dialect—on the same fixed skeleton.

The skeleton is built by iterative subdivision. Beginning from a bounding whole, the designer—or an algorithm—subdivides the form into successively finer substructures (for example 1 → 3 → 6 → 19), ensuring at each step that there are far more smalls than larges, that the ratio between consecutive levels stays within roughly two to six, and that the hierarchy is articulated through strong centers and well-defined boundaries (Alexander 2002–2005; Jiang 2015). A skeleton is treated as satisfying living-structure principles, and only then released for skin generation, once four checks are met: (i) the ratio of substructure counts between adjacent levels falls within the roughly two-to-six range specified by the scaling law; (ii) the L-score computed on the skeleton exceeds a pre-registered target set relative to the unrenovated or unrevised baseline; (iii) the B-score, computed independently from images of the skeleton via the Beautimeter, converges with the L-score's ranking of the design alternatives under consideration; and (iv) at least two independent raters agree, using the fifteen properties as a checklist, that strong centers and well-defined boundaries are present at each level. Because all four checks are computable or checklist-based rather than left to unstated intuition, the procedure can be reapplied by another research or teaching team to a different room, building, or design brief. The skeleton is not a style; it commits only to a particular kind of order. Once it is fixed, AI generates the skin—materials, ornament, surface articulation—in a contemporary Chinese, Beaux-Arts, modernist, or hybrid dialect, with several skins compared on cultural, climatic, or programmatic grounds. AI's role shifts decisively: it is no longer a generator of free-floating novelty but a structural mediator, a translator between universal order and particular expression. Stylistic differences are celebrated; structural mediocrity is not. The aim is structure before style, not structure instead of style. Nor is the paradigm a plea for historicism: radically contemporary skins can satisfy the criteria, and what it rejects is structural impoverishment, not modern materials or expression. The related worry that computable criteria could homogenize design, or be gamed, is taken seriously: because the skeleton constrains structure while leaving style free, diversity is built into the workflow rather than threatened by it.

The ordering of the terms is deliberate and substantive. Living structure is treated as more fundamental than AI for three reasons. First, ontologically, living structure is a property of the world—of rivers, trees, traditional towns, and the human body—discovered through centuries of observation rather than invented. A river goes on branching according to the same scaling, and a well-loved public square goes on holding people, whether or not any model exists to describe them. AI, by contrast, is a recent and changeable tool. To write “AI + Living Structure” would subordinate an enduring order of nature and culture to a particular technology. Second, logically, AI in this paradigm operates on criteria it does not itself supply: it generates and judges skins against the fifteen properties and the L- and B-scores, which come from living-structure theory. The theory furnishes the ends; AI furnishes a means. Third, pedagogically and ethically, placing living structure first keeps the human capacity to feel wholeness at the center, with AI in a supporting role. A paradigm that began with AI would risk training students to optimize whatever the model rewards; a paradigm that begins with living structure trains them to know what is worth optimizing for. The result is an approach that is neither anti-AI nor uncritically pro-AI, but structure-first: AI bound to scientific criteria of structural beauty, supporting but not replacing human judgement, and shifting the designer's role from form-maker toward guardian of living structure. This ordering also reframes what kind of science is at stake. Classical, mechanistic science prizes exactness—1 + 1 must equal 2—and is extraordinarily successful in engineering and physics. When applied uncritically to beauty and perception, however, this demand for certainty narrows rather than deepens our understanding. A twenty-first-century science of living structure should instead be statistical and tolerant of fuzziness: not pursuing perfect agreement but forming reproducible consensus within a good-enough range, much as large models do, also very much as stated by Henri Matisse (1947) ‘Exactitude is not truth’. Quantifying beauty is then like chasing a correlation coefficient that improves from 0.75 to 0.88 yet never reaches, and need not reach, 1.0. This fuzzy tolerance is not a compromise but a paradigm: it honors individual differences and sensory experience while keeping scientific explanatory power. It is, as Section 6 argues, exactly the disposition a liberal arts education should cultivate.

This ordering also marks a deliberate departure from the prevailing direction of educational technology, in which AI is placed first, and learning is reshaped around what the model can generate, score, or automate. In such an arrangement the criteria of value are quietly delegated to the system, and students learn, above all, to satisfy it. The place of generative AI in design studio pedagogy is itself an active and fast-growing debate, and this paradigm is offered as one position within it rather than as a departure from it. Recent systematic reviews report gains in creative output, conceptual variety, and design efficiency when AI is used in studio settings (Onatayo et al. 2024; Alamasi and Asfour 2026, a PRISMA-guided systematic review of generative AI in architectural design education published in Architecture), set against a recurring concern, echoed across Stanimirovic et al. (2026), a systematic review of generative AI as a pedagogical partner in conceptual design published in Buildings, that early or unscaffolded AI use can reduce students' reflective engagement even as it increases output. Observers of the design studio have likewise warned of this drift and called for explicit frameworks to govern how generative tools enter pedagogy (Iranmanesh and Lotfabadi 2025). The Living Structure + AI paradigm's insistence on a fixed, human-declared, checklist-verified skeleton before any AI-generated skin is produced is a direct, structure-first response to this second concern: it is designed so that reflective, criteria-setting engagement precedes rather than follows generative output. The Living Structure + AI paradigm reverses the dependency: the criteria come from a theory of what makes environments and lives whole, and AI is enlisted to help students meet them. The difference is not anti-technological but directional. It asks not how education should be remade to suit AI, but how AI can be bound to ends that education has always served—perception, judgement, and the felt sense of what is true and beautiful.

A final clarification is needed about the scope of this section's examples. The skeleton–skin workflow illustrated in Figure 4 is drawn from architecture and interior design because that is where the workflow is easiest to see and to measure; it is not offered here as a contribution to architectural or interior-design method to be judged on those disciplines' own terms. Rather, it is used in this paper as a concrete, hands-on exercise through which non-specialist students—most of whom, as Section 4 reports, have no architecture or design background at all—practise a more general liberal-arts habit: declaring explicit criteria before invoking a powerful generative tool, and keeping final judgement in human hands. The same two-stage logic—fix criteria and structure first, then let AI generate variable expression within that structure—is deliberately repeated in the non-architectural exercises described in Sections 5.2 and 5.3: the reflective AI-dialogue essays, and the self-as-living-structure reflection in which students first name their own 'skeleton' of values and habits before using AI-assisted dialogue to explore its expression. The architectural skeleton–skin example is thus one instance of a transferable pattern, not the paper's real subject.

## 4 Methods and Participants

This section consolidates, in one place, the research design, participants, measurement instruments, and data-collection procedures used across all six teaching settings reported in Section 5, and states plainly what kind of evidence each instrument produces. It replaces methodological detail that was previously scattered across the individual course descriptions.

### 4.1 Overall research design

The evidence reported in this paper comes from a teaching-embedded, design-based-research programme led by LivableCityLAB, a research group whose teaching-research project is run in partnership with Residential College 1 (RC1), whose student tutors assisted in delivering the Self and Wholeness sessions. The three strands reported in Section 5—a series of campus renovations embedded in credit-bearing courses, an undergraduate common-core course, and a postgraduate summer elective—are related but administratively separate offerings, run by the same lab and lead instructor, rather than a single, formally unified, university-wide curriculum; no claim is made in this paper about department- or university-wide adoption of the paradigm, or about consensus among faculty outside the reporting research group. Each of the four renovation cases, and each of the two courses, follows the same experiential-learning logic—design → build → experience → redesign → rebuild (Kolb 1984)—adapted to its own setting and timeframe. Data collection for the four renovation cases involved pre- and post-renovation measurement of the same physical space; data collection for the two courses involved rubric-scored, course-embedded assignments collected as part of normal instruction. All human-subject data collection was carried out under the institutional ethical approval reported in the Institutional Review Board Statement, with individual-level data shared only in aggregated, de-identified form.

### 4.2 Participants

Table 1 summarizes the term, enrollment, and composition of each setting reported in this paper. We emphasize, in light of the very different scale, duration, and toolchain of the undergraduate and postgraduate courses (Section 5.2 and 5.3; see also Appendix A2), that no data from the two cohorts are merged, aggregated, or jointly analyzed anywhere in this paper. The single simplified syllabus presented in Appendix A2 is a curricular design document—a shared backbone delivered at two distinct levels—not a shared analytic dataset, and results are reported separately for each course throughout.

Table 1: Participants, terms, and composition for the six teaching settings reported in this paper

| Setting | Term | N | Composition |
|---|---|---|---|
| Classroom W1-233 (city-science course) | Autumn 2023 | 25 (postgraduate) | Approximately equal gender split |
| Student Activity Centre 5A-220 (Living Structure-based Human-Centred Design in Campus Space Renovation, 8-week evening course) | Autumn 2024 | 8 (first-year Residential College students) | Approximately equal gender split |
| Office E3-312 and Meeting Room E3-314 | Evaluated as part of an MPhil thesis (Qian 2025) | 1 (MPhil) | See Qian (2025) for design-team composition |
| Self and Wholeness (Habits, Mindsets and Wellness, HMW, common-core course) | Spring 2026 | 49 (first-year undergraduates) | Scattered across UG programmes (e.g., AI, Data Science and Analytics, Smart Manufacturing); >90% report no architecture or design background |

| Experiential Learning: Living Structure + AI Inspired Design (UGOD 6100I, postgraduate summer elective) | Summer 2026 | 7 (2 MPhil/Master's, 5 PhD) | 3 female, 4 male; predominantly Urban Governance and Design Thrust, one Data Science and Analytics student |
|---|---|---|---|

**4.3 Measurement instruments**

Four instruments recur across the settings reported in Section 5. The L-score (Jiang and de Rijke 2023) is computed from an annotated image or plan of a space by counting substructures and their hierarchical levels (L = S × H); for the four renovation cases it was computed before and after the intervention from matched viewpoints. The B-score, produced by the Beautimeter (Jiang 2025), is computed by prompting a large vision-language model to rate images of a space against the fifteen properties; as noted in Section 2, because the Beautimeter is itself derived from the fifteen properties, agreement between the B-score and the L-score is treated in this paper as an internal consistency check between two related instruments, not as independent, theory-external validation.

The Mirror of the Self Test (MOST) (Alexander 2002c) is a paired-comparison preference instrument in which participants are shown before/after or alternative-design image pairs and asked which they recognize more of themselves in; we report the direction and consistency of preference, not a standardized psychometric score. 3M Visual Attention Software (3M-VAS) (Lavdas et al. 2021; Sussman and Hollander 2014) models predicted visual-attention heatmaps from images of a space, allowing comparison of attention concentration and pathing before and after an intervention.

Read together, these four instruments mirror the left-brain/right-brain duality set out in Section 2. The L-score and the B-score are rule-based, structural computations—an explicit, left-brain analysis of a space or image against the fifteen properties—and are best read as the pair’s rationality-facing instruments. MOST and 3M-VAS, by contrast, register human response directly, a stated preference and a predicted pattern of visual attention, and are best read as the pair’s sensibility-facing, right-brain instruments. Consistent with the “instruments of conversation, not oracles” stance taken in Section 2, none of the four is used, or should be read, as an absolute verdict that a design is good or bad in itself; all four are most meaningfully interpreted comparatively—between two images, two design alternatives, or the same space before and after an intervention—which is how they are applied throughout Section 5.

We state explicitly what these four instruments do and do not measure. All four are architectural and perceptual measures of a space or an image of a space; none is a standardized educational-psychology or well-being scale, and none was administered to students as a validated pre/post measure of learning, cognition, or well-being. Where the paper draws on these instruments (principally in Section 5.1), the resulting claims are about the measurable and perceived qualities of the renovated spaces themselves—a necessary but not sufficient condition for any broader claim about student learning. Separately, in Section 5.2 and 5.3, we report what the courses' own rubric-scored assignments, presentations, and declared Intended Learning Outcomes (ILOs) show about students' stated understanding and self-reported change; this is course-embedded, rubric-based formative evidence, not a validated standardized learning-outcomes instrument, and we describe it as such throughout.

**4.4 Data collection and analysis**

For the four renovation cases, the L-score, B-score, MOST, and 3M-VAS were each computed or compared once before and once after the physical intervention, from matched viewpoints and image sets; results are reported descriptively (direction and approximate magnitude of change) rather than through inferential hypothesis testing, given the small number of independently renovated spaces (four) involved. Across all four cases, the L-score increased after renovation, and the B-score approximately doubled; exact per-case figures are available from the author upon request, consistent with the Data Availability Statement. For the two courses, assessment data (participation records, written and hands-on assignments, presentations, and peer evaluations) were collected as part of normal, Institutional Review Board-approved instruction and scored against the rubrics reported in Appendix A2; these data are summarized narratively in Section 5.2 and 5.3 rather than subjected to statistical hypothesis testing. We flag this descriptive, non-inferential level of analysis explicitly as a limitation in Section 7, and identify the administration of a standardized pre/post educational-psychology instrument as a priority for the next phase of the research programme.

**5 Practices Led by LivableCityLAB: A Curriculum Built on Living Structure**

Theory must be tested in real space and real classrooms. Under its own teaching-research project, and in partnership with Residential College 1, LivableCityLAB has built a multi-level curriculum that uses living structure and the Living Structure + AI paradigm as its organizing spine, spanning public common-core courses, disciplinary courses, residential-college teaching projects, and public training. Three strands are described here: architecture and city-science courses, an undergraduate common core course, and a new postgraduate course. As Section 4 makes explicit, these are three related but administratively separate offerings run by the same lab and lead instructor, not a single, formally unified degree-level curriculum; a common logic runs through all three—"design → build → experience → redesign → rebuild" (an experiential-learning cycle in the sense of Kolb 1984)—so that structural reasoning and felt life keep correcting one another in real space rather than on screens or paper alone.

**5.1 Architecture and city-science courses: campus renovation as living laboratory**

The first strand uses the skeleton–skin workflow as the basis of teaching-led renovations on campus, within the same everyday-facility scope (classrooms, activity centres, offices, and meeting rooms) established in Section 1. Over three academic years, students—many without architectural backgrounds—studied four ordinary interiors: Classroom W1-233, Student Activity Centre 5A-220, Office E3-312, and Meeting Room E3-314. In each case students measured the room, declared a skeleton, evaluated AI-generated skins, followed the building by professional contractors, and then inhabited the result.

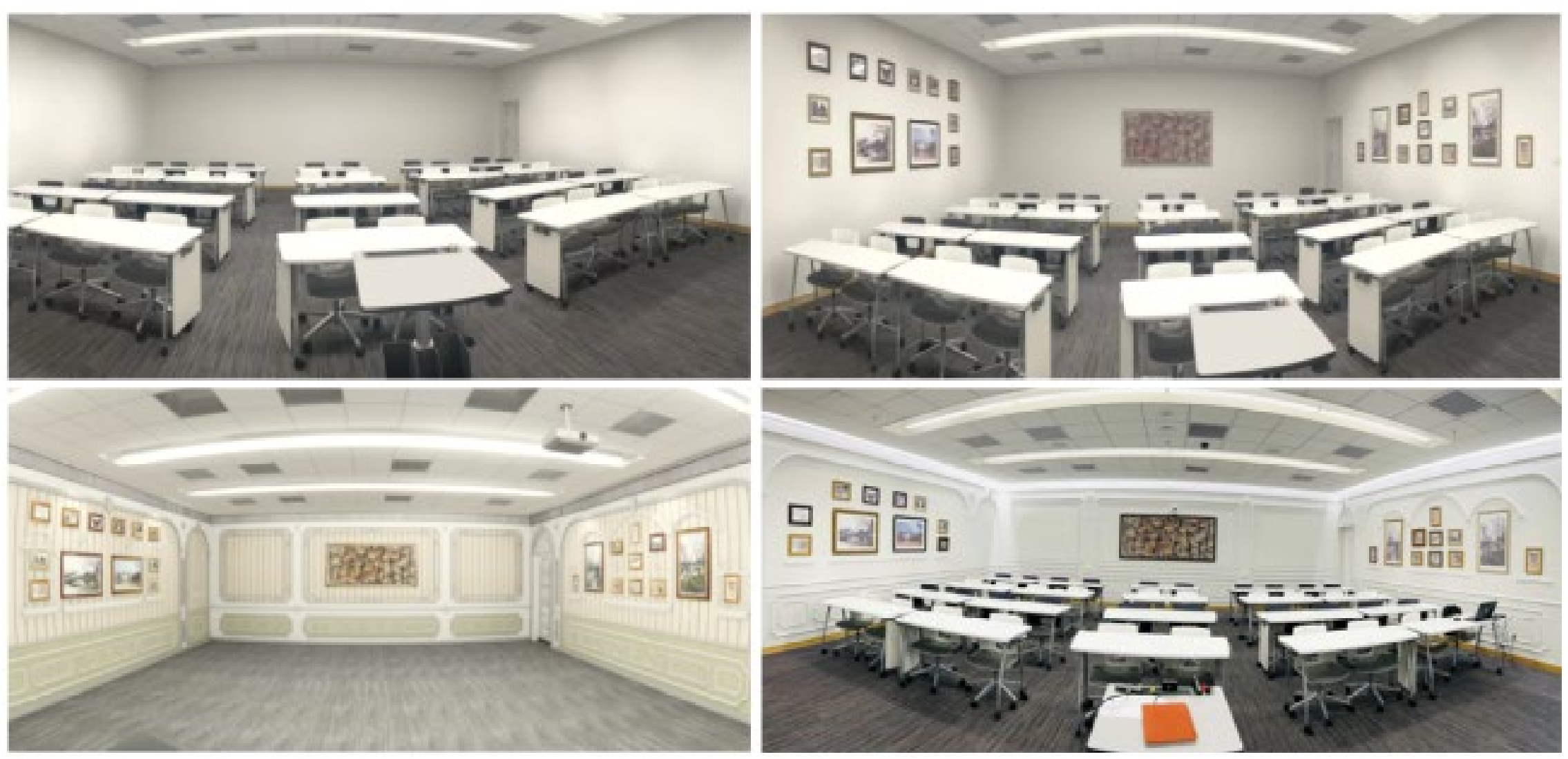

Figure 5: (Color online) Classroom W1-233 before (top/left) and after (bottom/right) the student-led renovation. The undifferentiated long wall is reorganized through paneling, moldings, and framed artworks that express a multi-level skeleton, while the room's size, furniture, and program are unchanged.

The Classroom W1-233 renovation (Figure 5) began within a city-science course, Autumn 2023, N = 25 postgraduate students, approximately equal gender split. The classroom, though used to teach a science of cities, was itself visually and spatially unscientific: an undifferentiated long wall with almost no legible order. In the opening class of the course, students were invited to treat their own classroom as a design problem, and to renovate it using the living-structure principles the course was about to teach. An undifferentiated long wall was reorganized into a skeleton of bays and panels across six levels of scale, raising the L-score by roughly an order of magnitude; the realized renovation introduced paneling, moldings, framed artworks, and articulated boundaries —twenty-four framed pictures and paintings, arranged across three levels of scale, were hung on the walls as part of the realized skeleton—and students overwhelmingly preferred the renovated condition. Across the cases, the L-score, the B-score, MOST preference tests, and 3M-VAS heatmaps converged: spaces with higher structural scores also felt more alive and drew more coherent attention (Sussman and Hollander 2014; Lavdas et al. 2021).

The Student Activity Centre 5A-220 renovation was carried out in a dedicated eight-week evening course, Living Structure-based Human-Centred Design in Campus Space Renovation, Autumn 2024, N = 8 first-year Residential College students, equal gender split. This multifunctional lounge and foyer, whose walls were flat and whose transitions were unmarked, saw student teams introduce nested centres, articulated boundaries, and fine-grained substructures while preserving the openness the programme required; the 3M-VAS heatmaps shifted from diffuse, weakly concentrated attention before the work to clear focal centres and continuous attention paths afterwards. Office E3-312 and Meeting Room E3-314 were subsequently renovated and evaluated as part of an MPhil thesis (Qian 2025), which carried the workflow into work-oriented rooms, relying on panelling, framing, mouldings, and lighting rather than structural change—interventions that were modest, reversible, and inexpensive. In every case the L-score and B-score rose—the B-score approximately doubling across all four cases—mirror-of-the-self pairs favoured the renovated condition, and these independent lines of evidence converged, reinforcing the lesson that structural metrics and felt judgements move together. As Section 4 notes, these results are architectural and perceptual measures of the spaces themselves, and the convergence between the B-score and the L-score in particular is treated as an internal consistency check between two related instruments rather than as independent validation of living-structure theory.

Three lessons stand out. The skeleton–skin workflow is teachable quickly even to non-architects; the renovations are governance-friendly—incremental, reversible, low-carbon, and respectful of existing footprints; and structural metrics and felt judgements reinforce one another. The work has been consolidated in a research monograph on the structural beauty of Chinese traditional architecture seen through AI (LivableCityLAB 2025) and disseminated through education-practice bases established with partners in China and abroad.

**5.2 Undergraduate liberal arts: Self and Wholeness**
The second strand brings living structure into general education. Self and Wholeness is a credit-bearing common core course within the Habits, Mindsets and Wellness (HMW) framework, delivered in Spring term 2026 to N = 49 first-year undergraduates drawn from across UG programmes, including AI, Data Science and Analytics, and Smart Manufacturing, among others. By the instructors' estimate, more than 90% of enrolled students had no prior architecture or design background at all; the course, and the renovation cases in Section 5.1, together provide evidence that living-structure principles are teachable to non-specialist students at every level of the university, from first-year undergraduates to PhD candidates. RC1 student tutors assisted in delivering the course. Through the lens of living structure, it invites students to rediscover wholeness—the deep coherence connecting the self, the built environment, and technology. Grounded in The Nature of Order (Alexander 2002-2005), the course combines lectures, hands-on making, field visits, dialogue, and reflective writing, pairing the analytical power of AI with the perceptual depth of human experience. Its intended outcomes range from diagnosing environments for "patterns of aliveness" to reflecting on the "self as a living structure," building sustainable habits, and using wholeness to strengthen teamwork.

The course unfolds over 13 weeks in three movements. The first, Having the Knowledge of Life, introduces the fifteen properties, the L-score and Beautimeter, MOST, and 3M-VAS scanning, with a perceptual-fieldwork visit to living urban fabric such as Yongqingfang or Dongchongli. The second, Creating Life in Our Daily Work and Life, is frankly hands-on: floral-art and handicraft workshops, column design, 3D printing, and AI-assisted wall design, so that students enact differentiation and adaptation at many scales. The third, Becoming a Life-Giving Person, turns inward—reading The Luminous Ground alongside Eastern philosophy, distinguishing the egocentric "self" from the deep, connected "I", and recognizing the maker's shift from imposer of will to channel for wholeness, in which beauty is the visible expression of truth. AI is used throughout under a clear policy of declared, acknowledged use—a declaration-and-acknowledgment statement is required on every assessed task, specifying which generative AI tool was used, for what purpose, and accompanied by a kept record of prompts and responses (see Appendix A.2 for the full policy)—including AI-assisted dialogues on themes such as rationality versus sensibility and mechanism versus organism. The aim is not information but transformation: inner coherence, sharpened perception, and the capacity to create beauty in everyday surroundings.

Self and Wholeness runs across a semester, and its rhythm is as much social and embodied as it is intellectual. The first movement takes students on perceptual fieldwork to living urban fabric such as Yongqing Fang or Dongchongli, where they compile a livingness analysis with the Beautimeter, and it includes a guest lecture on new traditional architecture. The second fills the studio with making—floral art, handicraft, column design carried through to 3D printing, and AI-assisted wall design—each exercise a small experiment in differentiation and adaptation at a different scale. The third turns reflective and even playful, pairing a close reading of The Luminous Ground and Eastern philosophy with a comedy workshop on self and wholeness, and culminating in a presentation, "My Journey Toward Wholeness," in which students account not only for their finished work but for how the process changed their self-awareness. Assessment is weighted accordingly: participation 15%, written and hands-on assignments 20%, presentation 20%, and peer evaluation 5%, embedded in the year-long HMW common-core framework, rewarding participation, hands-on assignments, that final presentation, and peer evaluation rather than examination.

As Section 4 notes, the evidence available from this course is course-embedded, rubric-based formative evidence—drawn from the participation records, written and hands-on assignments, and rubric-scored presentations described above and in Appendix A.2—rather than a validated, standardized measure of educational or psychological outcomes. Within that scope, instructors report that most students, the great majority of whom began the course with no design background, were able by its end to discuss the wholeness or beauty of a built environment in the course's vocabulary, to connect it to accounts of well-being and healing effects of living spaces, and to produce simple, people-friendly re-designs; we offer this as a course-embedded, formative observation rather than as a demonstrated, generalizable learning-outcomes finding.

**5.3 Postgraduate course: Experiential Learning — Living Structure + AI Inspired Design**

The third strand carries the approach to the postgraduate level. Experiential Learning: Living Structure + AI Inspired Design (course code UGOD 6100I; three credits) is a summer-school course that mirrors the three-part rhythm of Self and Wholeness but raises the technical and reflective demands. It was delivered in Summer term 2026 to N = 7 postgraduate students (2 MPhil/Master's and 5 PhD students; 3 female, 4 male), predominantly from the Urban Governance and Design Thrust, with one student from Data Science and Analytics; as this course was still in progress at the time of writing, the figures reported here reflect its completed portion. Students study the fifteen properties and the L-score, judge architecture for well-being with AI tools, take MOST, and work through an experiential-learning tour of the renovated campus spaces and nearby living fabric, comparing contemporary and traditional environments on and off campus. They then move through architectural-component design, 3D printing, laser cutting, 3D scanning and modelling, and column design and construction, before presenting not only finished work but their personal growth—their journey of connecting with the "I" and the luminous ground. As at undergraduate level, generative AI is used openly, under the same declaration-and-acknowledgment policy, as a structural mediator and cognitive amplifier rather than as a substitute for the student's own perception and making.

\Its declared outcomes ask students not only to diagnose environments for patterns of aliveness and to shape them through hands-on making, but also to reflect on the self as a living structure that grows through awareness and care, and to integrate emotional, intellectual, and spiritual insight for balanced personal development. Assessment is weighted as participation and engagement 20%, weekly assignments 50%, and presentation 30%—substantially different from the undergraduate weighting given the course's smaller cohort, longer contact hours (3 hours/week versus 1 hour 20 minutes/week), and fuller fabrication toolchain (laser cutting and 3D scanning are added at this level). The same design → build → experience → redesign rhythm governs the work, so that rising technical sophistication— laser cutting, 3D scanning and modelling, fabricated columns—never drifts away from felt life.

**6 Implications: Liberal Arts Education Down to Earth**

The central implication for liberal arts education is that living structure provides a concrete way to develop the whole person. Because the same phenomenon can be felt and computed, a single object of study—a wall, a column, a street, a flower arrangement—exercises both hemispheres at once: the left brain measures levels of scale and L-scores, the right brain registers whether the result feels alive. This is the paper's central pedagogical aim rather than a claim already demonstrated by the evidence reported in Sections 4 and 5, which is architectural, perceptual, and course-embedded rather than psychometric; whether the pedagogy in fact reunites the two hemispheres in any lasting or measurable sense is a question for the longitudinal follow-up work described in Section 7. Whole-person and liberal arts education, at bottom, attempt the same reconciliation that the fuzzy, living science of Section 3 describes: developing rational thought while reawakening sensory perception, so that both hemispheres come truly alive. Living structure gives that abstract aspiration a daily, doable form.

A second implication concerns the place of AI in education. The courses treat AI as a cognitive amplifier: it helps students articulate why a space feels alive, scores candidate designs against shared criteria, and generates skins to compare—while the structural decisions, the felt judgements, and the making remain the student's own. This is a deliberate counter to the drift toward AI as a content machine. By binding AI to living-structure criteria, the curriculum models a disposition students can carry into any domain: use the tool to extend perception and judgement, not to outsource them.

A third implication is for disciplinary education. The same paradigm that grounds the liberal arts strand reorganises architecture and city-science teaching around reading built environments structurally, declaring and verifying a skeleton before generating a skin, and using AI in dialogue with the L- and B-scores. It scales upward from the classroom to the city: urban informatics has long described and predicted cities in detail (Batty 2013) but said less about what cities ought to be; living structure offers one route from “is” to “ought,” because an environment whose livingness can be measured can also be improved (Jiang 2025). Because most spaces that need to become more alive already exist, the paradigm proposes a third path between leaving buildings untouched and demolishing them: incremental, reversible, structurally guided enrichment (Mehaffy and Salingaros 2015).

Taken together, these implications describe a particular stance toward living with AI, and thus a particular kind of citizen that the curriculum aims to cultivate. A student trained to declare criteria before invoking a tool, to ask whether a generated result is genuinely more alive rather than merely more novel, and to keep final judgement in human hands carries a transferable disposition into a world saturated with automated content. This is AI literacy of a deeper kind—not fluency in prompting, but the habit of subordinating powerful means to humane ends. We advance this as the paper's normative argument for why this pedagogy is worth pursuing, not as an empirical finding about its effects: in an era when the danger is less that machines will think than that people will gradually stop noticing what matters, an education that keeps perception, feeling, and embodied making at its centre is, we argue, a quiet form of resistance—one that any discipline, not only design, can adopt, and one whose actual long-term effects remain to be tested.

The deepest implication is methodological and ethical. Both the renovations and the courses insist that learning moves from the virtual to the real, from the screen to lived space, privileging human-centered design and the participant's own embodied, intuitive experience. Embodiment—physical models, 3D printing, wall and column making, floral art, fieldwork—is not decoration around a digital core but the core itself, the means by which structural reason and felt life keep correcting one another. This is the sense in which the work aims to design or produce plans not only on paper and on screens but, as the phrase has it, down to earth—in the daily life spaces students can touch and remake. It is also a shift from cold technology toward warm intelligence, in which coherence, warmth, and felt life are treated as primary objectives rather than residual outputs. The measure of success is not how efficiently a space or a tool performs, but whether the people who live with it feel more whole; efficiency is welcome, but it is not the point. A liberal arts education conducted this way aims to produce graduates who are at once technically capable and humane, analytical and sensitive, rooted in tradition and open to the future—graduates the curriculum is designed to help recognize and create wholeness in the world and in themselves, though confirming that this aim is achieved is a task for the longitudinal research described in Section 7, not a claim made on the strength of the present evidence.

**7 Conclusion and Future Work**

This paper argued that living structure offers a timely foundation for liberal arts education in the era of AI. As a phenomenon that is both felt and computed, it aims to reunite the two hemispheres of the brain; as the basis of a Living Structure + AI paradigm—deliberately so ordered, because structure is more fundamental than the tool that expresses it—it puts AI in the service of human perception rather than the reverse; and as the spine of a curriculum spanning architecture and city-science courses, an undergraduate whole-person education course, and a new postgraduate course, it has shown that abstract theory can be turned into perceivable, experiential, and verifiable practice in real space. Beauty, as Alexander wrote, is not about how something looks but about how it is; this paradigm is, in the end, an argument for taking that sentence seriously in education. In Cooley's (1987) terms, it is an argument for educating architects rather than bees—people who carry the criteria of wholeness in imagination before any machine builds.

The evidence reported here is early, and its limits should be stated plainly. It is architectural, perceptual, and course-embedded rather than psychometric: the L-score, B-score, MOST, and 3M-VAS results reported in Section 5.1 measure the renovated spaces themselves, not students' cognition or well-being directly, and no standardized pre/post educational-psychology scale was administered in any of the settings reported here; the course-embedded assessment evidence reported in Section 5.2 and 5.3 is rubric-based and formative rather than a validated learning-outcomes instrument. It comes largely from a single institution, from a specific research group (LivableCityLAB) rather than the university as a whole; preference tests, attention maps, and structural scores are partial proxies for well-being rather than direct measures of it; course cohorts have so far been modest in size (N ranging from 7 to 49 across the settings reported here), and no data from different cohorts are pooled; convergence between the B-score and the L-score reflects internal consistency between two related instruments rather than independent, theory-external validation (Section 4); and the cultural transferability of both the criteria and the pedagogy remains an open empirical question. These limitations define the research programme rather than undermine it.

Several lines of work follow. The teaching-research project is committed to a three-to-four-year programme of repeated teaching implementation and case accumulation, through which the curriculum's content and methods can be systematically evaluated and refined—especially their flexibility, scalability, and transferability across course types and spatial scales. On the technical side, promising directions include subdivision algorithms that generate skeletons from boundary conditions, AI judges that build on the Beautimeter and large-model studies, a genuinely independent, theory-free measure of perceived aliveness that could move the B-score's relationship to living-structure theory from an internal consistency check toward external validation, and the calibration of a "moderate" rather than maximal level of livingness. The administration of a standardized pre/post educational-psychology or well-being instrument, alongside the architectural and course-embedded measures already in use, is a priority for the next phase of this programme. On the educational side, the most important task is longitudinal: to study how an education grounded in living structure shapes students' perception, well-being, and capacity to make their surroundings more alive over time. The wager of this work is that, embedded in a paradigm that takes structure seriously, AI can help more students, in more cultures, learn to see and create wholeness—on the earth, and in themselves.

In the immediate term, the two new courses provide a natural testbed: their first full cohorts can be followed with the same instruments used on the renovated rooms—L- and B-scores, MOST, and reflective writing—so that changes in how students perceive and make space can be tracked rather than merely asserted. Beyond LivableCityLAB's own setting, the curriculum is designed to be adopted and adapted elsewhere, and we invite colleagues in other institutions and cultures to test whether an education built on living structure travels, and what it must become in order to do so.

**Data Availability Statement**
The data supporting the findings of this study are openly available for scrutiny. The two course syllabuses that underpin the teaching practice—the undergraduate common-core course Self and Wholeness and the postgraduate course Experiential Learning: Living Structure + AI Inspired Design—share a common curricular backbone and are therefore presented together as a single simplified reference syllabus in Appendix A.2, with the two delivery levels, and the ways in which they differ (contact hours, toolchain, and assessment weighting; see Section 4.2 and the summary at the start of Appendix A.2), indicated throughout, so that it can be readily adopted or adapted for other teaching settings and levels. The four campus renovation case studies discussed in the paper are documented as interactive before-and-after immersive renderings, whose links are also listed in Appendix A, with each link opening an environment in which readers can switch between the pre- and post-renovation scenes and inspect the structural and perceptual changes for themselves. The structural and perceptual measures reported in the paper—L-scores, B-scores (Beautimeter), 2–6 scaling-ratio checks, visual-attention (3M-VAS) heatmaps, and the mirror-of-the-self preference judgments—were derived from these materials and are available from the corresponding author upon reasonable request. Because the human-subject responses were collected under the institutional ethical approval reported in the Institutional Review Board Statement, any individual-level data are shared only in aggregated, de-identified form, consistent with the conditions of that approval and with participants' informed consent.

**Acknowledgements**
An earlier draft of this paper was prepared with the assistance of a large language model for language refinement and structural organization. The author takes full responsibility for the content and any remaining errors. The author is grateful to the two anonymous reviewers for their highly constructive comments, which significantly improved the quality of the paper. Thanks are also extended to current and former students and colleagues at LivableCityLAB and the Urban Governance and Design Thrust, as well as to colleagues at Residential College 1 (RC1), HKUST(GZ).

This work was supported by several research grants, including the HKUST(GZ) Research on Practices (ROP) projects for 2025 and 2026; the AI Research and Learning Base of Urban Culture under Project No. 2023WZJD008; the Guangdong Provincial Key Laboratory of Integrated Communication, Sensing and Internet of Things under Grant No. 2023B1212010007; The Hong Kong University of Science and Technology (Guangzhou) under Grant No. G0101000142; and the City–University Joint Fund of the Guangzhou Science and Technology Project under Grant No. 2024A03J0529.

## Appendix A: Supporting Materials

This appendix gathers the materials referenced in the Data Availability Statement: the interactive renderings of the four campus renovation case studies (Appendix A1) and a single simplified reference syllabus (Appendix A2). Because the undergraduate and postgraduate courses share a common curricular backbone but differ substantially in contact hours, toolchain, and assessment weighting (summarized in Table A1 below), only one syllabus is given; it retains the backbone of the courses as actually taught, condensed so that other educators can adopt or adapt it, and it indicates the two levels at which the course is delivered and where those levels diverge. Both courses were initially inspired by the Self and Wholeness course developed within the Building Beauty program in Italy, and were subsequently re-designed for the undergraduate and postgraduate settings at HKUST(GZ) with the addition of the Living Structure + AI toolchain (the Beautimeter and L-/B-scoring, the 3M visual-attention scan, and AI-assisted generative design).

Table A1: Summary of how the undergraduate and postgraduate delivery levels of the shared syllabus differ. No data from the two courses are pooled; see Section 4.2.

| Dimension | Undergraduate (Self and Wholeness) | Postgraduate (Experiential Learning) |
|---|---|---|
| Contact hours | 1 hour 20 minutes / week | 3 hours / week |
| N (this paper) | 49 | 7 |
| Toolchain | Beautimeter, L-/B-scoring, 3M-VAS, AI-assisted design | Adds laser cutting and 3D scanning/modelling |
| Assessment weighting | Participation 15% + assignments 20% + presentation 20% + peer evaluation 5% | Participation and engagement 20% + assignments 50% + presentation 30% |

### Appendix A1: Campus Renovation Case Studies (Interactive Before/After Renderings)

Each link opens an interactive environment in which the viewer can switch between the pre-renovation and post-renovation scenes and navigate the space from multiple viewpoints, making the skeleton–skin transformation and the associated gains in L-score and B-score directly perceptible (Table A2).

Table A2: Virtual reality (VR) tours of selected teaching, activity, office, and meeting spaces at HKUST(GZ)

| Space | URLs |
|---|---|
| Classroom W1-233 | https://vr.justeasy.cn/view/174v3j47i0470778-1734793858.html |
| Student Activity Center 5A-220 | https://vr.justeasy.cn/view/147e374o54n19057-1753083138.html |
| Office E3-312 | https://vr.justeasy.cn/view/zn61187413306813-1756382846.html |

| Meeting Room E3-314 | https://vr.justeasy.cn/view/uk17d141l920b639-1758796620.html |
|---|---|

**Appendix A2: Simplified Reference Syllabus**
This single syllabus underlies both courses described in the paper—the undergraduate Self and Wholeness and the postgraduate Experiential Learning: Living Structure + AI Inspired Design. Because the two share a common backbone, only one is given here; it can be delivered at either level, and Table A1 above summarizes where the two levels diverge.

**Undergraduate delivery:** 1 hour 20 minutes per week, as a Whole-Person-Education unit of the common-core course Habits, Mindsets and Wellness (HMW). N = 49 in the reported cohort (Spring 2026).
**Postgraduate delivery:** 3 hours per week, as an elective (course code UGOD 6100I; 3 credits). N = 7 in the reported cohort (Summer 2026).
**Duration:** 13–14 weeks in one semester, organised in three parts; the longer postgraduate sessions accommodate the fuller fabrication toolchain.
**Medium of instruction:** English.
**Aim.** To let students rediscover life and wholeness — the coherence connecting the self, the built environment, and technology—through the lens of living structure, combining the analytical power of AI with the perceptual depth of human experience.
**Intended learning outcomes.** By the end of the course, students will be able to:
Demonstrate an understanding of relationship between environmental quality ("wholeness," "living structure") and its impact on holistic well-being.
Demonstrate a capacity to diagnose environments (natural, built, social) for "patterns of aliveness" and apply this awareness to shape one's surroundings through hands-on making.
Demonstrate self-awareness and a capacity to self-manage, set and pursue goals, by reflecting on the "self as a living structure", and learning to respond to challenges and failure as a natural part of the creative process.
Build sustainable habits that promote personal and interpersonal development by cultivating a continuous awareness of the patterns of "aliveness" in one's immediate environment and daily routines.
Utilize the principles of wholeness to enhance team effectiveness in diverse settings, harnessing different strengths to create coherent and life-enhancing collective work.
Weekly schedule. The thirteen weekly sessions are organised into three parts, whose activities are summarised below.

**Part 1 — Having the knowledge of life: the phenomena of life (Weeks 1–5)**
Students are introduced to the course and to Christopher Alexander's life work, the definition of life and wholeness, and a scientific way to measure beauty and to design. Across this part they study the 15 properties of living structure in built forms; practise L-scoring and the Beautimeter; take the Mirror-of-the-Self Test (MOST) and the visual-attention scan (3M-VAS); work through an architectural-design case study with spatial-design hands-on; judge architecture for well-being with an AI tool; attend a guest lecture and critical dialogue on traditional versus contemporary architecture; and undertake a site visit with perceptual fieldwork, reported as a group study of the livingness of a chosen place (Beautimeter required). An in-class quiz closes the part.

**Part 2 — Creating life in daily work and life: design practice and technological fusion (Weeks 6–10)**
Students move from analysis to making. Activities include a floral-arts workshop and a handicraft workshop based on living structure; a column-design exercise carried from concept to a digital model and then to 3D-printing hands-on (with laser cutting added in the postgraduate sessions); AI-assisted design based on living structure, applied to interior-wall design; and 3D scanning and modelling of architectural space in the postgraduate sessions. Toward the end of this part, students begin a reflective essay written in dialogue with AI.

**Part 3 — Becoming a life-giving person: wholeness, cognition, and the inner turn (Weeks 11–13)**

The focus turns inward. Students present their work and join a reflective workshop on self and wholeness; read deeply into The Luminous Ground and the unified field, distinguishing the egocentric "self" from the connected "I"; reflect on the transformation of the maker and the unity of beauty and truth, analysing their own design process and revising the essay; and give a final reflection and presentation, "My Path of Creation," covering the product, the process, and their personal growth.

Teaching approach. Experiential (walks, sketches, sensory exercises), hands-on (design at several scales), dialogical (reflective dialogue and circle sharing), integrative (bridging science, art, and wholeness), and embodied (mindfulness and presence). The same Living Structure + AI toolchain is used at both levels: the Beautimeter and L-/B-scoring, the 3M visual-attention scan, and AI-assisted generative design; the longer postgraduate sessions add laser cutting and 3D scanning.

Assessment. Participation and engagement, weekly hands-on assignments, and a final presentation, with peer evaluation added at undergraduate level. Indicative weightings—undergraduate: participation 15%, written and hands-on assignments 20%, presentation 20%, peer evaluation 5%, embedded in the year-long HMW common-core framework; postgraduate: participation and engagement 20%, weekly assignments 50%, final presentation 30%. Generative-AI use in assessed tasks is permitted under a declaration-and-acknowledgment policy aligned with the university's academic-integrity guidelines.